\documentstyle[epsf,titlepage,fleqn,12pt]{article}

\begin{document}
\baselineskip=20.5pt
\parskip=3pt

\begin{titlepage}
\title{Diffusion in disordered lattices and related Heisenberg ferromagnets}
\author{M. D. Grynberg$^1$ and R. B. Stinchcombe$^2$\\
\\
1- Departamento de F\'{\i}sica, Universidad Nacional de La Plata,\\
C.C. 67, (1900) La Plata, Argentina\\
\\
2- Theoretical Physics, Department of Physics,
University of Oxford\\
1 Keble Road, Oxford OX1 3NP, UK}
\date{}

\maketitle

\begin{abstract}
We study the diffusion of classical hard-core particles
in disordered lattices within the formalism of a
quantum spin representation. This analogy enables
an exact treatment of non-instantaneous correlation
functions at finite particle densities in terms
of {\it single} spin excitations in disordered ferromagnetic
backgrounds. Applications to diluted chains and 
percolation clusters are discussed. It is found that density fluctuations
in the former exhibit a stretched exponential decay while an anomalous
power law asymptotic decay is conjectured for the latter.

\end{abstract}

\end{titlepage}

\section{Introduction}

Processes involving classical particles constrained to diffuse
and interact stochastically on discrete substrates is one
of the oldest schemes for the study of a rich variety of relaxation phenomena
\cite{Bouchaud,Havlin}. Although from a conceptual standpoint such processes 
give only a phenomenological level of understanding, they do provide the type
of behavior necessary to describe irreversibility in many-body systems 
which  otherwise, would be difficult to derive from first principles.
Among the main techniques to obtain the time probability distribution of 
these processes, possibly the master equation (MEQ) approach \cite{vanKampen} 
is the more directly related to physical concepts. Despite its apparent 
simplicity however, it generally gives rise to an infinite hierarchy of coupled
equations whose solutions become quite involved  to elucidate, particularly  
at large times.
In a concerted effort to remedy this situation, quantum field theory 
methods have recently regained new impetus in the study of the dynamics of
{\it non}-quantum-mechanical-objects \cite{Mattis}.
Basically, the underlying idea is that  the MEQ
resembles a time-dependent Schr\"odinger equation in a pure imaginary
time with probability distributions playing the role of wave functions.
Thus, by studying the field theory associated to the MEQ a formal solution,
in some cases {\it exact}, of many-body
probability distributions can be found explicitly \cite{we1}.

As a contribution to the current momentum of this approach,
in this work we shall revisit the problem of hard core particles diffusing
in quenched disordered lattices emphasizing the formal 
analogy with  its quantum counterpart.
The avoidance of double occupancy accounts for the essential
aspects of both inter-particle interactions and mobility whereas frozen-in 
disordered bonds $\{J_{\bf r, r'}\}$, 
modeling random hopping rates between locations $\{{\bf r,r'}\}$,
is essential to understand the slowing down of transport properties in a
vast family of inhomogeneous and glassy systems \cite{Bouchaud,Havlin, Alex1}.

Following the thread of ideas initiated in \cite{Alex2} and developed
subsequently by many authors \cite{Mattis,Gwa}
here we attempt to further advance the subject forward by means {\rm of
a (pseudo) spin-$\frac{1}{2}$ description in which spin up or 
down at a given site
corresponds to particle or vacancy, say, at that location.
The MEQ is then equivalent to the action of a quantum spin
"Hamiltonian"  encompassing the original transition  probability rates.
Interestingly, for the case of symmetric diffusion of hard core particles
the resulting Hamiltonian reduces to an  isotropic 
Heisenberg ferromagnet
whose full rotational symmetry applies {\it irrespective} of the 
hopping disorder.
Although this continuous symmetry is ultimately imposed by
conservation of probability
throughout  the Brownian process, its exploitation is not evident
without appealing to the quantum spin Hamiltonian analogy \cite{we2}. 
Particularly, this formalism becomes advantageous in analyzing 
spontaneous fluctuations  of the steady state such as 
non-instantaneous density-density  correlation functions
(e.g. structure factors and scattering functions).
As we shall see, the use of elementary quantum
mechanical considerations along with selection rules based on the 
conservation of total angular momentum enable an exact treatment 
of these {\it many}-body correlations
in terms of {\it single} spin excitations.

The layout of the paper is organized as follows. In Section 2,
we recast the MEQ of the disordered Brownian process as a physical 
realization of the  Heisenberg ferromagnet.
Section 2 A and 2 B treat in turn the structure of the steady
state and its non-instantaneous two-point correlations.
Section 3 discusses autocorrelation functions in a variety
of scenarios. An exact treatment of these functions in diluted chains 
is given in Section 3 A, where a stretched exponential asymptotic dynamics  
exhibiting a scaling regime is found. Section 3 B continues with a
discussion of diluted lattices in higher dimensions. Exploiting
scaling pictures for the frequency and the density of
states of single-excitations on percolation clusters 
{\it at criticality} \cite{Aharony}, we are led to suggest 
a slower or anomalous diffusive kinetics.
We end the paper with Section 4 which contains our conclusions,
along with some remarks on extensions of the present work.

\section{Master equation and spin representation}

A central assumption underlying most phenomenological stochastic models is that
the actual non-equilibrium dynamics of real systems can be 
approximated by a discrete
Markovian process and, therefore described by a MEQ.
The latter governs entirely the time evolution of the
probabilities $P(s,t)$ of finding the system in a certain
configuration $\vert s \rangle$ at time $t$.  If $W(s \to s')\,$
denotes the rate or transition probability per unit time at which
configuration $\vert s \rangle\,$ evolves
to $\vert s' \rangle\,$, the MEQ reads
\begin{equation}
\label{master}
\partial_t \,P(s,t) = \sum_{s'} \left[\: W(s'\to s)\, P(s',t)\,
- \, W(s \to s') \,P(s,t)\:\right]\,.
\end{equation}
Assuming the basis vectors $\vert\,s\,\rangle\,$  form an
orthonormal set,
it is useful to describe the ensemble averaged state vector of the
system at time $t$ as 
$\vert\,P(t)\,\rangle =  \sum_s\, P(s,t)\,\vert\,s\,\rangle$. Thus,
starting from a given probability distribution
$\vert \,P (0) \, \rangle$ it can be readily checked that 
the formal integration of Eq.\,(\ref{master}) yields
a state vector solution of the type
\begin{equation}
\vert\,P(t)\,\rangle = e^{\,-H\,t }\, \vert\,P(0)\,\rangle \,,
\end{equation}
where the matrix elements of the transition operator $H$
(or "Hamiltonian") are constructed as \cite{Kawasaki}
\begin{eqnarray}
\label{non-diag}
\langle\,s'\,\vert\,H\,\vert\,s\,
\rangle &=& -\,W(s\to s')
\hspace{0.4cm},\hspace{0.4cm} s' \ne s\,,\\
\label{conservation}
\langle\,s\,\vert\,H\,\vert\,s\,\rangle &=&
\sum_{s'\ne s}\, W(s \to s')\,.
\end{eqnarray}

The steady states of our stochastic processes correspond to
the ground states of $H$ of each subspace within which
the dynamics takes place, all with the zero eigenvalue.
Any eigenvalues with positive real parts $\lambda > 0\,$
correspond to decaying states with lifetime $1/\lambda\,$.
Due to probability conservation, clearly every column of
$H$ adds up to zero, i.e. $\sum_{s'}\,\langle s' \vert H
\vert s \rangle = 0\,$.
Thus, in passing and for future reference
it is worth pointing out that a {\it left} steady state
can be immediately obtained as
\begin{equation}
\label{left}
\langle\, \psi^{^{^{\!\!\!\!\sim}}}\,\vert =
\sum_s\, \langle\, s \,\vert\,,
\end{equation}
which evidently satisfies $\langle \psi^{^{^{\!\!\!\!\sim}}}\,\vert H = 0$.
Also, notice that
$\langle \psi^{^{^{\!\!\!\!\sim}}}\,\vert P(t) \rangle \equiv 1\; \forall t$.

Turning to the analogy between diffusion of hard-core classical
particles and quantum spin-1/2 systems we now represent a particle
or vacancy, respectively, at site ${\bf r}$ by spin up or down,
i.e. by $m_{\bf r} = +1, -1$ where $m_{\bf r}$ is an eigenvalue
of the $z$-component (say) of the Pauli operator 
$\vec \sigma ({\bf r})$ for site ${\bf r}$.
Clearly a typical configuration
$\vert s\rangle$ in a general lattice with $N$ locations
can be characterized by a direct product
$\vert \,m_{ {\bf r}_1},\, ... \,, m_{{\bf r}_N}\,\rangle$ 
of all the site spinors.
So the set $\{W ( s \to s')\} = \{J_{\bf r\, r'}\}$ of transition hopping
rates between connected  sites ${\bf r, r'}$ (e.g. nearest-neighbors
$\langle {\bf r, r'} \rangle\,$),
then describes the following spin-exchange process
$$\left\vert\, m_{{\bf r}_1} \; ... \; m_{\bf r}=1 \; ...\;
 m_{\bf r'}=-1\;...\; m_{{\bf r}_N}\rangle\; \right.
\longleftrightarrow^{^{^{\!\!\!\!\!\!\!\!\!\!\!\!\!\!\!J_{_{\bf r\, r'}}}}}\;
\left\vert\, m_{{\bf r}_1} \; ... \; m_{\bf r}=-1 \; ...\;
 m_{\bf r'}=1\;...\; m_{{\bf r}_N}\rangle\; \right.\,.$$

Here and in the following the set $\{J_{\bf r\, r'}\}$ can be arbitrarily
substitutionally disordered.
To construct the associated "Hamiltonian" $H$ whose terms
keep proper track of both
probability conservation and the spin exchange process, it is useful to
cast the discussion in terms of spin raising and lowering
operators $\sigma^+_{\bf r}, \sigma^-_{\bf r}$.
The off-diagonal part of $H$ exchanging configurations
as schematized above, is therefore given by
\begin{equation}
\label{off}
\sum_s\, \sum_{s' \ne s}\, \vert s' \rangle\, \langle s'
\vert H \vert s \rangle\,\langle s \vert =
-\, \sum_{\langle \bf r\,r'\rangle} \,
J_{\bf r\,r'}\left ( \, \sigma^+_{\bf r} \,\sigma^-_{\bf r'} \,+\,
{\rm h.c.}\,\right)\,,
\end{equation}
whereas conservation of probability, i.e. Eq.(\ref{conservation}),
requires the emergence of a diagonal part of the form
\begin{equation}
\label{diagonal}
\sum_s  \vert s \rangle \,\langle s \vert H \vert s \rangle\,
\langle s \vert =
\sum_{\langle \bf r \, r' \rangle}\,J_{\bf r\,r'}
\left [\; \hat n_{\bf r}\, (1\,-\,\hat n_{\bf r'})\, +\,
(1\,-\,\hat n_{\bf r})\,  \hat n_{\bf r'}\; \right]\,,
\end{equation}
where $\hat n_{\bf r} \equiv \sigma^+_{\bf r}\,\sigma^-_{\bf r}$
denote occupation number operators.
These terms basically count the total number of ways in which a
given configuration $\vert s \rangle$ can evolve to different states
$\vert s' \rangle$ through a particle hopping attempt between nearest
neighbor pairs $\langle \bf r \, r' \rangle$, weighting each accessible
attempt with rate $J_{\bf r\,r'}$. This yields precisely the right
hand side of Eq.\,(\ref{diagonal}).
Thus, in terms of usual spin-1/2 Pauli matrices
$\vec \sigma \equiv (\sigma^x, \sigma^y, \sigma^z)$,
by virtue of Eqs. (\ref{off}) and (\ref{diagonal})
the evolution operator $H$ reduces finally to a Heisenberg ferromagnet
\begin{equation}
\label{Heis}
H = -\,\frac{1}{2} \sum_{\langle \bf r\,r'\rangle} \,J_{\bf r\,r'}\,
 \left (\,\vec \sigma_{\bf r}\,\cdot \vec \sigma_{\bf r'}\,-\,1\,\right)\,.
\end{equation}
Interestingly, due the isotropic nature of the interactions,
the stochastic dynamics leaves invariant all the components of the total 
angular momentum
${\bf S} = \frac{1}{2} \sum_{\bf r} \vec\sigma_{\bf r}$,
namely $[\,H , {\bf S}\,] = 0\,$
{\it irrespective} of the disordered background of hopping
rates $J_{\bf r\,r'}$. As a consequence of the full rotational symmetry,
the calculation of spontaneous density fluctuations 
are simplified remarkably (Section 2 B).  
In preparation for the analysis of those functions, let us first examine 
the form of the steady state at finite particle densities.

\begin{center}
{\bf A. The steady state}
\end{center}

Evidently, the fully jammed  ferromagnetic state  
$\vert\, F\,\rangle = \vert\,S\,,S^z\,\rangle$
with  total spin $S = N/2$ and total magnetization $S^z = N/2$
is a steady state (SS).  Since $[\, H\,,\,S^-\,]=0\,$
we can generate  normalized SS
$\vert \,\psi_m \,\rangle$ with $\vert\,S\,,S^z\,\rangle = 
\vert N/2, \,N/2-m\,\rangle$, i.e. having  $m$-vacancies
and particle density $\rho = 1 - m/N$, and such that 
$H \,\vert \,\psi_m \,\rangle = 0\,$ by applying $m$-times 
the lowering operator  $S^- = \sum_{\bf r}\,
\sigma^-_{\bf r} = S^x - i\,S^y\,$, namely
\begin{equation}
\label{steady}
\vert\,\psi_m\,\rangle = \alpha_m \: (S^-)^m\,\vert\,F\,\rangle\,,
\end{equation}
where the normalization factor is $\alpha_m = \sqrt{(N-m)!/N!\,}$.

Notice that $\vert\psi_m\rangle\,$ is an equally weighted
linear combination of all permissible $\Omega = {N \choose m}$
configurations with $(N-m)$ particles. In particular,
\begin{equation}
\label{coincide}
\langle\,\psi_m \,\vert = \frac{1}{\sqrt{\Omega}}\,
\langle\, \psi^{^{^{\!\!\!\!\sim}}}\,\vert\,,
\end{equation}
which is in line with the more basic observation that,
aside from normalization prefactors, left and right SS should coincide
since Eq\,(\ref{Heis}) involves a self-adjoint evolution operator, namely
$W ( s \to s') = W (s' \to s)$.
From this latter observation, it is worth pointing out that
detailed balance arises immediately in
Eq.\,(\ref{master}) given that $P(s) \to 1/\Omega\;\;
\forall \;\; \vert s\rangle$.

The structure of these equilibrium states is rather trivial
regardless of the hopping disorder. This follows by noting that
\begin{equation}
\hspace{-0.5cm}
\hat n_{\bf p}\,\hat n_{\bf q}\,\vert\,\psi_m\,\rangle =
\frac{1}{\sqrt{\Omega}\,}\: \sum_{j_1 <}\!"\,...\,
\sum_{\!\!\!\!\!...< j_m}\!"\:
\sigma^-_{{\bf r}_{j1}}\,...\,
\sigma^-_{{\bf r}_{jm}}\,\vert\,F\,\rangle\,,
\end{equation}
where the index ordering denotes sums over
all reachable different configurations
and the double prime restricts the sums to ${\bf r}_j \ne {\bf p, q}$.
This yields ${N-2 \choose m}$ non-vanishing configurations and therefore
\begin{equation}
\langle\, \psi_m\,\vert \,\hat n_{\bf p}\,\hat n_{\bf q}\,
\vert\,\psi_m\,\rangle =
\frac{{N-2\choose m}}{\Omega} = \left(1-\frac{m}{N}\right)
\left(1-\frac{m}{N-1}\right)\;\; \sim_{_{_{\!\!\!\!\!\!\!\!\!\!\!\!
N \to \infty}}} \, \rho^2 \,,
\end{equation}
so density-density correlators decouple in the large system limit. 
A  similar reasoning holds for many-point correlators thus yielding
spatially uncorrelated SS. Henceforth, we 
address our approach to the correlations between
fluctuations that occur spontaneously at {\it different} times
which as is known  \cite{Kubo}, are closely related to the 
relaxation dynamics governing non-equilibrium regimes.

\begin{center}
{\bf B. Non instantaneous steady state correlations}
\end{center}

We will be especially interested in the calculation of {\it non-instantaneous}
joint probability distributions $\langle\,{\cal A }(t)\,{\cal B}(0)\,\rangle$
of quantities $\cal A, B$ such as local densities or local density
correlations, i.e. diagonal operators in the particle
or $\sigma^z$ representation. Here, the brackets indicate
an average over histories up to a to time $t$ starting from
a given probability distribution $\vert P(0) \rangle$. More precisely,
this can be expressed in terms of the following discrete path integration 
\begin{eqnarray}
\nonumber
\langle\,{\cal A }(t)\,{\cal B}(0)\,
\rangle &=& (\Delta t)^{n-1}\,\sum_{s_1}\,...\,\sum_{s_n}\,
P(s_1,0)\,\langle \, s_1\, \vert {\cal B }\,\vert s_1\, \rangle\,
 W (s_1 \to s_2)\,\\
\label{history}
&\times&\, W (s_2 \to s_3)\;\;...\;\; W (s_{n-1} \to s_n) \,
 \langle\, s_n\, \vert {\cal A }\,\vert\, s_n\, \rangle\,,
\end{eqnarray}
where the sums run over all possible states
$\vert s_j\rangle$, and $\Delta t \,W (s_{j-1} \to s_j)$
denotes the probability of evolving from
$\vert s_{j-1} \rangle $ to $\vert s_j \rangle$
in a single elementary time step $\Delta t = t/(n-1)$ \cite{steps}.
However, by construction [Eqs. (\ref{non-diag}) and (\ref{conservation})\,],
this latter probability is given by
$\langle\, s_j\,\vert\,1 - \Delta t\, H\,\vert\,s_{j-1}\,\rangle\,$.
Because $\cal A, B$ are taken diagonal in the particle
representation, 
\begin{eqnarray}
\nonumber
P(s,0)\, \langle\,s\,\vert\,{\cal B}\, \vert\,s\,\rangle &=&
\langle \,s\,\vert \,{\cal B}\,\vert\,P(0)\,\rangle\,,\\
\langle\, s \,\vert \,{\cal A }\,\vert\, s\, \rangle &=&
\langle\, \psi^{^{^{\!\!\!\!\sim}}}\,\vert\,{\cal A}\, \vert\,s\,\rangle\,,
\end{eqnarray}
so, recalling that $(1- \Delta t\, H)^n \to e^{-H\,t}$, 
in the limit of a large number of steps \cite{steps}, 
Eq.\,(\ref{history}) yields
\begin{equation}
\label{functions}
\langle\,{\cal A }(t)\,{\cal B}(0)\,\rangle =
\langle \psi^{^{^{\!\!\!\!\sim}}}\,
\vert\, {\cal A }\,e^{-H\,t}\,{\cal B }\,\vert P(0)\,\rangle\,.
\end{equation}
Since $[\,H , {\bf S}\,] = 0$ for any disorder realization, 
Eq.\,(\ref{functions}) allows for a systematic treatment of
spontaneous density fluctuations. In terms of the local density 
fields referred to above, this requires consideration of
non instantaneous particle-particle  correlations $C_{\bf r,r'}(t) = 
\langle \,\hat n_{{\bf r}}(t)\, \hat n_{\bf r'}(0)\, \rangle$ 
occurring in the SS distribution  $\vert\, P(0)\,\rangle = 
\frac{\vert\,\psi_m\,\rangle}{\sqrt{\Omega\,}}$ with $\rho = 1-m/N\,$.
As usual, it is convenient to choose a basis  
$\{ \vert \Lambda \rangle = \vert \,S\,, S^z\,, 
\lambda\,\rangle\}$ of common eigenstates of ${\bf S}^2\,, S^z$ and
$H$ with eigenvalues $S (S+1)\,, S^z$ and $\lambda$ respectively.
Inserting  an orthogonalized set of
$\vert \Lambda \rangle$  states in Eq.\,(\ref{functions}), 
by virtue of Eq.\,(\ref{coincide}) it follows that in the 
$m$-down spin subspace  $C_{\bf r,r'}(t)$ can be written as
\begin{equation}
\label{correlation}
C_{\bf r,r'}(t) = \sum_{ {\rm levels}\;\, \lambda\, >\, 0}\,
e^{-\,\lambda\,t}\,\langle\,\psi_m\,\vert\,\hat n_{{\bf r}}\,
\vert\,\Lambda\,
\rangle\,\langle\,\Lambda\,\vert\,
\hat n_{\bf r'}\,\vert\,\psi_m\,\rangle\,,
\end{equation}
where we have restricted the sum to eigenstates
$\vert \Lambda\rangle\,$ different from  $\vert\,\psi_m\,\rangle\,$
(which has vanishing eigenvalue $\lambda = 0$),
to subtract the time independent correlations, i.e. 
$\langle \hat n_{{\bf r}} \rangle 
\langle \hat n_{{\bf r'}} \rangle = \rho^2\,$.
All eigenvalues $\lambda$  are real and positive definite, as by construction
$H$ is a self adjoint stochastic operator.

Selection rules can now be applied to simplify remarkably 
the calculation. Noting that the Pauli matrices
$\sigma^z_{{\bf r}}\,$ are tensors of rank one, since
$\hat n_{{\bf r}} = (1 + \sigma_{{\bf r}}^z)/2\,$,
the Wigner-Eckart theorem
ensures non-vanishing matrix elements
$\langle \psi_m \vert \,\hat n_{{\bf r}} \,\vert
\Lambda\rangle\,$ only if the total spin $S\,$ of
$\vert \psi_m \rangle\,$ and $\vert \Lambda \rangle\,$
differ by 0 or 1, i.\,e. $S=N/2,\, N/2-1\,$; in either
case with $S^z = N/2 -m\,$. To identify the relevant 
$\vert\Lambda\rangle\,$ states, we first compute the total spin $S\,$ 
of a {\it single} spin excitation
$\vert\,\varphi_{\lambda}\,\rangle = \sum_{{\bf r}}\,
\varphi_{\lambda}({\bf  r})\:\sigma^-_{{\bf r}}\,\vert\,F\,\rangle\,$.
Since ${\bf S}^2 = S^z\,S^z\,+\,\frac{1}{2}\,
(\,S^+\,S^- + S^-\,S^+\,)\,$, it can readily checked that
\begin{equation}
\label{spin}
{\bf S}^2\,\vert\,\varphi_{\lambda}\,\rangle = 
\cases{ (\frac{N}{2}-1)\,\frac{N}{2}\,\vert\,
\varphi_{\lambda}\,\rangle \:\: {\rm if} \:\: \lambda > 0 \,,\cr \cr
\frac{N}{2}\,(\frac{N}{2}+1)\,\vert\,
\varphi_{\lambda}\,\rangle  \:\: {\rm if} \:\:  \lambda = 0\,,}
\end{equation}
for which the wave function should satisfy
$\sum_{{\bf r}} \varphi_{\lambda}({\bf r}) \equiv
\sqrt{N} \,\delta_{\lambda,0}\,$.
Therefore, the wanted  $\vert \Lambda \rangle\,$ states
giving non-zero matrix elements in Eq.\,(\ref{correlation}) are 
essentially rotated versions of the above single
spin excitation states. Specifically, recalling that the angular momentum
algebra imposes
\begin{equation}
S^-\,\vert\,S,\,S^z\,\rangle =
\sqrt{(S+S^z)\,(S-S^z+1)\,}\,
\left\vert\,S,\,S^z\!-\!1\,\right\rangle\,,
\end{equation}
we can generate a  normalized SS in the $m\,$- down spin sector
by applying $(m-1)\,$-times the lowering operator $S^-\,$ to the
spin excitation $\vert\,\varphi_{\lambda}\,\rangle\,$, namely
\begin{equation}
\label{Goldstone}
\vert\,{\Lambda}\,\rangle = \beta_m\, (S^-)^{m-1}\,
\vert\,\varphi_{\lambda}\,\rangle\,,
\end{equation}
where the normalization factor arising from the products
generated by each application of $S^-\,$ turns out to be
\begin{equation}
\beta_m =  \sqrt{\,\frac{(N-m-1)!}{(N-2)!\,(m-1)!}}\:\:,\:\:
1\, <\, m\, <\, N-1\,.
\end{equation}
Thus, for $\vert \Lambda \rangle \ne \vert \psi_m \rangle\,$
these are {\it all }  the linearly independent states contributing to 
Eq.\,(\ref{correlation}), having total spin $S= (N/2 - 1)\,$,
total magnetization $S^z = (N/2 - m)\,$, and $\lambda > 0$.
So, we are now left with the calculation
of $\langle\,\psi_m\,\vert\,\hat n_{{\bf r}}\,\vert\,
\Lambda\,\rangle\,$. This matrix element is expanded as
\begin{eqnarray}
\nonumber
\langle\,\psi_m\,\vert\,\hat n_{{\bf r}}\,\vert\,
\Lambda\,\rangle &=& \beta_m\,\langle\,\psi_m\,\vert\,
\hat n_{{\bf r}}\,(S^-)^{m-1}\,\vert\,
\varphi_{\lambda}\,
\rangle\,\\
&=& \beta_m\,\sum_{\bf p}\,
\varphi_{\lambda}({\bf p})\,
\langle\,\psi_m\,\vert\,
\hat n_{{\bf r}}\,(S^-)^{m-1}\,
\sigma^-_{\bf p}\,\vert\,F\,\rangle\,.
\end{eqnarray}
To go a further step in our analysis notice that
\begin{eqnarray}
\nonumber
\hat n_{{\bf r}}\,
(S^-)^{m-1}\,\sigma_{\bf p}^-\,\vert
\,F\,\rangle &=& (m-1)! \:
(1 - \delta_{\bf r,p})\\
&\times&
\sum_{j_1 <}\!"\sum_{j_2 <}\!"...\!\!
\sum_{\!\!\! ... < j_{m-1}}\!\!\!\!\!"\:\:
\sigma^-_{{\bf r}_{j_1}}\,...
\sigma^-_{{\bf r}_{j_{m-1}} }\,
\sigma^-_{\bf p}\: \vert\,F\,\rangle\,,
\end{eqnarray}
where the double prime restricts the
sums to vectors ${\bf r_j \ne r, p}\,$. 
Thus, for ${\bf r \ne p}\,$ there are {\Large
${N-2 \choose m-1}\,$} terms contributing equally to
$\langle\,\psi_m\,\vert\,\hat n_{{\bf r}}\,\vert\,
\Lambda\,\rangle\,$. Hence, by taking into
account the wave function constraint
$\sum_{{\bf r}} \varphi_{\lambda}({\bf r}) \equiv 0\,$
so as to ensure a total spin $S=N/2-1$\, ($\lambda > 0$),
we obtain
\begin{eqnarray}
\nonumber
\langle\,\psi_m\,\vert\,\hat n_{{\bf r}}\,\vert\,
\Lambda\,\rangle &=& (m-1)!\,m!\,{N-2 \choose m-1}\,
\alpha_m\:\beta_m\,\sum_{\bf p}\,\varphi_{\lambda}({\bf p})\,
(1-\delta_{\bf r,p})\,\\
&=& -\,\sqrt{ \frac{m\, (N-m)}{N\,(N-1)\,}}\:
\varphi_{\lambda}({\bf r})\,,
\end{eqnarray}
where $\alpha_m$ is taken as in Eq.\,(\ref{steady}).
Therefore the particle-particle correlations of 
Eq.\,(\ref{correlation}) are finally given by 
\begin{equation}
\label{main}
C_{\bf r,r'}(t) = \frac{N\, \rho (1-\rho)}{N-1}\!\!
\sum_{{\rm single \,\, \lambda\,\,levels} }\!\!\!\! e^{\,-\lambda\, t}\,
\varphi_{\lambda}({\bf r})\,
\varphi^*_{\lambda}({\bf r'})\,,
\end{equation}
which at most requires the evaluation of {\it single} spin-wave 
(Bloch) excitations in disordered ferromagnetic backgrounds.
This is the main result  of this section. 
Certainly, the usefulness of the quantum spin operational
formalism developed so far is subject to the knowledge
of such excitations.  However, much is known about their 
properties and density of states \cite{Aharony,Houches}, and in fact
this enables us to obtain explicit results, particularly for large 
time asymptotic regimes (Section 3). 
A similar reasoning for non-instantaneous 
$\nu$-point correlators would involve $\vert \Lambda \rangle$ states 
with total spin $S = N/2 - {\cal M}\,$, i.e. states with 
${\cal M}$-{\it interacting} magnons (${\cal M}=1,\,...\,,\nu\,$). 
In this more general case however, analytical progress seems difficult 
given the scarcity of exact results in $d > 1$, even
for ordered substrates \cite{Wortis}.

\section {Autocorrelation functions of diluted lattices}

A subcase of Eq.\,({\ref{main}) entailing particular interest      
and which is related to the probability of hard core random walkers
returning to the origin, is that of autocorrelation functions $C(t)$.     
This corresponds to averaging ({\ref{main}) over all locations ${\bf r = r'}$.
For isotropic structures, it is plausible that the number of sites
a random walker (particle) has visited after a time $t$ is proportional 
to the volume $R^d(t)$, where $R$ is the root mean square 
displacement in a substrate of dimensionality $d$ 
(either Euclidean or fractal).  Thus for  densities $\rho \ll 1/R^d$\,, 
the probability of returning to the origin should in principle 
scale as $C(t) \propto R^{-d}(t)$. Departures from normal or Fickian 
diffusion ($ R\propto t^{1/2}$\,) in disordered lattices can therefore 
be studied by both, i) evaluating the asymptotic behavior of 
autocorrelations and , ii) averaging the hopping distribution.

Autocorrelations are in turn closely related
to the density of states in the substrate. In fact, due to 
the normalization of the single level eigenfunctions,  
after  averaging over the sites (origins) of a given sample, 
in the large $N$-limit clearly Eq.\,(\ref{main}) reduces to 
\begin{equation}
\label{auto}
C (t)  = \frac{ A_{\rho}}{N}\!\! \sum_{{\rm single \,\,
\lambda\,\,levels} }
\!\!\!\! e^{\,-\lambda\, t}\,,
\end{equation}
where $A_{\rho} \equiv \rho (1-\rho)\,$.
So, the form of the density of states for $\lambda \to 0$
determines entirely the large time asymptotic behavior of
the autocorrelation function for hard core particles at finite densities.

Before continuing and in preparation for the analysis 
of autocorrelations in disordered scenarios, it is worthwhile to pause 
and  consider the finite size scaling regimes emerging asymptotically 
from Eq.\,({\ref{auto})  in regular {\it ordered} situations
($\{J_{\bf \langle\, r\, r' \rangle }\} \equiv J\,$).
Let us consider briefly a $d\,$-dimensional
hypercubic lattice  with $N = L^d$ sites and periodic
boundary conditions. It is well known that
the corresponding $\lambda$ levels are $2\,J\,\sum_{j=1}^d \,
(1 - \cos \frac{2 \pi n_j}{L})\,$
with $n_j = 0,\,\cdots L-1$. From Eq.\,(\ref{auto}) it is
straightforward to check that the autocorrelation
function of such finite  system factorizes as
\begin{equation}
\label{factors}
C(t) = A_{\rho}\,\frac{e^{-2 d J t}}{L^d}\,\left[\:\left(\:
\sum_{n=0}^{L-1}\, e^{\:2 J \,t\, \cos \,
\frac{2 \pi n}{L} }\:\right)^d\,-\:1\:\right]\,,
\end{equation}
where the last term takes care of the
cancellation of the $\lambda = 0$ contribution.
Thus, it can be readily verified that there is a scaling regime 
$L \to \infty,\, J t \to \infty\,$ for which
the autocorrelation scales with a universal function 
${\cal F}(\tau)$ so long as $\tau \equiv  J t/L^2\,$ 
is held constant, namely
\begin{eqnarray}
\nonumber
C(L, t) &=& (J t)^{-d/2}\: {\cal F} (\tau)\,,\\
\label{scaling0}
{\cal F} ( \tau) &=& A_{\rho}\,\tau^{d/2}\,
\left[\,\vartheta_3^{^{d}}\,(q)\,- \,1 \right]\;,\;\;
\vartheta_3\, (q) \equiv 1 \, + \, 2\,\sum_{n=1}^{\infty}\, q^{n^2}\,,
\end{eqnarray}
where $q =e^{-4 \pi^2 \tau}$ and $\vartheta_3\, (q)\,$ is a Jacobi theta
function of the third kind \cite{Abram}\,.
Hence, the typical size of the system sets the time scales 
($t \propto L^2$\,) for which the dynamics becomes diffusive. 
Also, recalling that $\lim_{_{_{_{\!\!\!\!\!\!\!\!\!\!\!\! \tau \to 0}}}}
 \sqrt \tau \: \vartheta_3\, (q) = \frac{1}{2 \sqrt\pi}$,
the autocorrelations of the infinite system result in long 
time diffusive tails (as they should), whose asymptotic
kinetics is given by
\begin{equation}
\label{asympt}
C(t) \sim \frac{A_{\rho}}{(4 \pi)^{d/2}} \: (J t)^{-d/2}\,.
\end{equation}

Turning to more general situations, a case of interest 
is that of bond-(hopping-) diluted  lattices. More specifically, 
we will focus attention on 
the following binary hopping probability distribution
\begin{equation}
\label{binary}
P(J') = p\, \delta_{J,J'} \,+\, (1-p)\, \delta_{J',0}\,,
\end{equation}
defined uniformly throughout all lattice bonds. For short-range hoppings,
say e.g. nearest-neighbors, a purely geometric effect,
known as the percolation transition, arises from the configurational
aspects caused by dilution. We address the reader to Ref. \cite{Stauffer} 
for a detailed introduction to this subject which has been found useful
to characterize a rich variety of diluted systems including spin 
systems \cite{Stinchcombe}.
Here, we just mention that there exists a critical concentration 
$p_c$ below which only finite clusters exist and 
above which a cluster spanning the (infinite) lattice is formed.

Of course in $d=1$ any bond removal disconnects the lattice, so
$p_c = 1\,$. However, as we shall see in Section 3 A, 
as in the regular case [\,Eq.\,(\ref{scaling0})\,],
autocorrelations also exhibit an asymptotic 
scaling regime for {\it finite} dilutions  but through a 
dilution-dependent scale of length.
In higher Euclidean dimensions, extensive research 
\cite{Aharony,Houches,Stauffer,Mandelbrot} 
has led to the conclusion that percolation clusters are statistically
self-similar at the transition, and so are random fractals.
They are characterized by various scaling dimensions (fractal, spectral)
for the various processes (mass, density of states) measurable on the 
fractal. Thus, exploiting scaling results for the spectral dimension,
in Section 3B we shall discuss the 
implications which the resulting density of states imposes on
the large time average behavior of Eq.\,(\ref{auto}).

\begin{center}
{\bf A. Diluted chain}
\end{center}

In the one-dimensional case for $0< p < 1$ the distribution (\ref{binary})
breaks the $N$-site chains into
a collection of finite {\it disconnected} segments,
each having a  number $1 \le L \le N-1$
of consecutive bonds, or $L+1$ correlated sites.
So,  the probability $W (L)$ to find a chain with
$L$ consecutive non-vanishing transition
rates and  free boundaries is independent of
the chain location and is given by
\begin{equation}
\label{distri}
W (L) = \cases{ (1-p)^2 \,p^L  \:\: {\rm if} \:\: 1 \le L \le N-2\,,\cr \cr
 (1-p) \, p^L  \:\:\: {\rm if} \,\:\:\: L = N-1\,,\cr \cr
 p^L  \:\:\: {\rm if} \,\:\:\: L = N\;\;{\rm (periodic\;\; chain)}.}
\end{equation}
On the other hand,  as shown in Section 2, 
for a given segment of $(L-1)$ bonds,
it is sufficient to consider the corresponding evolution operator $H_L$
within the {\it single} spin excitation sector.
From Eq.\,(\ref{Heis}), it can be readily checked that  $H_L$
can be recast in terms of the following $L \times L$ tridiagonal form
\begin{equation}
H_L = 2\,J\, \left[\begin{array}{ccccc}
1 & -1 &\;\; 0 & \cdots &0\\
-1  & 2 & -1  & \ddots &\vdots\\
0 & \ddots & \ddots & \ddots& 0\\
\vdots & \ddots &-1 & \;\;2 & -1 \\
0 & \cdots &\;\; 0 & -1 & \;\;1
\end{array}\right]\,,
\end{equation}
with eigenvalues
\begin{equation}
\label{eigen}
\lambda_L (n) = 4\,J\,\left( 1- \cos \frac{n \pi}{L}\,\right)
\,,\;\;\; n = 0\,,1\,, \cdots \,,L-1\,.
\end{equation}
Assuming a constant particle density $\rho$ per chain,
the overall autocorrelations must be averaged over {\it both}
histories (time) and samples (hopping disorder).
Thus, the autocorrelations arising
from all possible distributions of disconnected segments reads
\begin{equation}
\label{auto0}
\langle C(t) \rangle = A_{\rho}\,\sum_{L=1}^{N-1} \,
W(L)\,\sum_{n=1}^L \,e^{-\lambda_{L+1} (n) t}\,,
\end{equation}
where $A_{\rho}$ is taken as in Eq.\,(\ref{auto}).
Here the first sum, denoted by the brackets on the left hand side,
runs over segments of $L$ consecutive bonds and takes into account
the hopping disorder, whereas the second sum runs over eigenvalue levels
$\lambda_{L+1} (n) > 0,\:\: 1 \le n \le L$, and carries out the
time dependent average.
We are especially interested to elucidate the asymptotic behavior
of Eq.\,(\ref{auto0}) for {\it arbitrarily} large times and
{\it finite} dilution regimes $0 < p < 1\,$. To this aim, it is useful
to introduce the natural length scale emerging in the problem, namely
the percolation correlation length
\begin{equation}
\xi = -\frac{1}{\ln p}\,,\:\:\: 0 < p < 1\,,
\end{equation}
which measures the effective distance between missing hopping rates.
On general grounds, this scale
can be regarded as the mean distance between two sites belonging
to the same cluster which, in higher dimensions, diverges 
as $\xi \sim (p-p_c)^{-\nu}\,$
with a universal exponent $\nu$ depending solely on the space 
dimensionality \cite{Stauffer}.

Now, from Eqs.\,(\ref{distri}) and (\ref{auto0})
it follows  that in the thermodynamic limit $N \to \infty$
we are left with the calculation of
\begin{equation}
\label{auto2}
\langle C(t) \rangle = A_{\rho}\,\frac{(1-p)^2}{p}\,
\sum_{L=2}^{\infty}\; \sum_{n=1}^L\,
 e^{-\left [\frac{L}{\xi}\, +\,\lambda_L (n)\,t \right]}\;.
\end{equation}
Since $\lambda_L (n) = 2J (n \pi   /L)^2 + {\cal O} (1/L^4)$
it is clear that the dominant contributions to ({\ref{auto2})
when $t \to \infty$, are basically  contained in large $L$ segments.
Introducing the (dimensionless) scaling variables
\begin{equation}
\label{scale}
\tau = \frac{J t}{\xi^2}\,,\:\:\:s = \frac{L}{\xi}\,,
\end{equation}
it is straightforward to verify that  within the scaling regime
$\xi \gg 1$ with $J t \gg 1$ Eq.\,(\ref{auto2}) can be written as
\begin{equation}
\langle C(\tau) \rangle \sim \frac{A_{\rho}}{\xi}\,\int\limits_0^{\infty}\,
 e^{-s} \,\sum_{n=1}^{s \,\xi}\,e^{-\,2 \pi^2 n^2 \tau/s^2}\, d\,s\,.
\end{equation}
For $\tau \ll 1$, the sum over $n$ can be replaced by an integral,
and a result $\langle C(t) \rangle \propto t^{-1/2}$ (of pure-system form)
follows. This is the asymptotic behavior on one side of a crossover
occurring at $\tau \sim 1\,$. The other side has the more interesting
(disorder - dominated) asymptotic behavior, obtained by considering
$\tau \gg 1\,$. Then the sum is dominated by the $n=1$ term, so
\begin{equation}
\label{auto3}
\langle C(\tau) \rangle \sim \frac{A_{\rho}}{\xi}
\,\int\limits_0^{\infty}\,
e^{ -\,(s \,+\,\frac{2 \pi^2}{s^2}\tau)}\, ds\,.
\end{equation}
To obtain the asymptotic behavior of the integral 
we use a saddle-point expansion around the minimum  of the exponent, 
at $s_0 = (4 \pi^2 \tau)^{1/3}$, 
which becomes exact in the scaling limit $\tau \gg 1$.
After elementary manipulations, this finally yields 
autocorrelations characterized by a universal scaling function
${\cal U}(\tau)$ exhibiting a
stretched exponential decay together with a subdominant power law prefactor.
More explicitly, for $\tau \gg 1\,$, $\xi \gg 1\,$,
\begin{eqnarray}
\nonumber
\langle C(\xi\,,\tau) \rangle &=& \xi^{^{-1}}\,{\cal U}(\tau)\,,\\
\label{scaling}
{\cal U}(\tau) &=& A_{\rho} \,\sqrt {\frac{2 a \pi}{3}} \: \tau^{1/6}\:
\exp\,\left(\, -\,\frac{3 a }{2}\, \tau^{1/3}\, \right)\,,
\end{eqnarray}
where $a = (2 \pi)^{2/3}$. This is the central result of this sub-Section,
and it corresponds to localization when the average cluster size is smaller
than the pure diffusion length.

To check the reliability of this form we address
the reader's attention to Fig. 1 in which autocorrelations
are computed directly from Eq.\,(\ref{auto2}). The data
collapse obtained for different dilutions clearly confirm 
Eq.\,(\ref{scaling}) and provides further evidence
for the existence of an asymptotic scaling regime characterized
by the variable $\tau = J\,t/\xi^2\,$ and showing the crossover
at $\tau \sim 1\,$ between pure system diffusive behavior 
($\tau < 1\,$) and localization ($\tau > 1\,$).
Notice that so long as $p \ne 1\,$, no matter how small the dilution is,  
it induces at sufficiently large time a different dynamics 
from that developed by the regular, purely diffusive case 
[\,Eq.\,(\ref{asympt})\,]. This contrasting behavior comes
from the interplay of a pure diffusion length ($\propto t^{1/2}\,$)
and the percolation length $\xi\,$ (roughly the average segment length)
which diverges at the threshold $p=1\,$.

\begin{center}
{\bf B. Percolation cluster}
\end{center}

In higher dimensions the dynamic behavior is much richer
by virtue of\,  i) the more interesting fractal structure 
of the percolation geometry at the threshold $p_c\,$,
and\, ii) the existence of the infinite (spanning) cluster
above $p_c\,$. We therefore have to distinguish finite
cluster contributions to $\langle C(t) \rangle\,$
from the infinite cluster contribution when $p \ge p_c\,$.

We first consider how such features might affect the Goldstone
argument \cite{Nambu}. The spin symmetry properties remain
as for the pure system. The full rotational symmetry by an angle 
$\alpha$ around an arbitrary spin direction $\hat n$,
evidently leaves the evolution operator (\ref{Heis})
invariant, i.\,e.\,  
$H =  e^{- \,i \, \alpha\,{\bf S} \,\cdot \,\hat n}\, H\, 
 e^{\,i \, \alpha\,{\bf S} \,\cdot \,\hat n}\,$.
However, the steady (ground) states of Eq.\,(\ref{steady})
do not preserve such symmetry $\forall \,\hat n \,\ne \,\hat z$.
Because of this spontaneous symmetry breaking,
in the thermodynamic limit $N \to \infty$\,  
a low-lying band of gapless or Goldstone modes can be expected 
in the pure system \cite{Nambu} and, because of its unlimited size,
also on the infinite cluster ($p \ge p_c\,$)
irrespective of the hopping disorder in $H$.
Moreover, the Goldstone modes are  the $\vert\, \Lambda\,\rangle$ states
already introduced in Eq.\,(\ref{Goldstone}) and those with large
characteristic scale are ultimately
responsible for the emergence of a slow asymptotic kinetics 
[\,Eqs.\,(\ref{factors}),(\ref{scaling0})\,]\,.

Next consider the finite clusters (for $p \ne p_c\,$).
They are distributed with size distribution having
characteristic scale $\xi$ but nevertheless extending
with exponential tail to infinite size (see also Section 4).
Without these tails the finite size cutoff would introduce
a gap in the spectrum of the (Goldstone) modes on the ensemble 
of finite clusters.

The contributions, to dynamic properties such as 
$\langle C(\xi, t) \rangle\,$, from excitations on finite
clusters and on the infinite cluster (for the case $p \ge p_c\,$)
can be obtained from the density of single levels
$\omega (\lambda)$ or density of states (DOS) involved in
Eq.\,(\ref{auto}).  Therefore, the averaged autocorrelations 
can be written as 
\begin{equation}
\label{auto4}
\langle C(t) \rangle =  A_{\rho}\, \int\limits_0^{\infty}\,
\omega (\lambda) \, e^{ -\lambda\,t}\, d \lambda\,.
\end{equation}

The specific form of the DOS  
for $\lambda \to 0\,$ from modes on the finite clusters
is closely related to the role played by 
Lifshitz tails $\sim e^{-\lambda^{-r}}$ ($ r > 0\,$)
in the energy distribution,
a typical issue occurring in the presence of disorder \cite{Lifshitz}.
Consider, for example, a square lattice of which a fraction
$p$ of bonds are accessible. Below the percolation threshold
the low energy DOS is characterized by Lifshitz tails
which are contributed by large regions of connected sites. 
In the one dimensional case (Section 3 A) strings of length $L$ occur with 
probability $p^L$ and contribute low-lying ferromagnetic modes
with $\lambda \propto L^{-2}$, so the DOS form is 
$\omega (\lambda) \propto  p^{\lambda^{-1/2}}$.
In fact, such a tail in the DOS is dominant 
in determining the  relaxation dynamics
referred to in Section 3 A.

Above the percolation threshold Lifshitz tail modes are also present, 
but are swamped by the power law tail $\omega (\lambda) \sim \lambda^{d/2-1}$ 
due to low energy  excitations on the infinite cluster. Thus for $p > p_c$
we expect the usual diffusive decay  $\langle C (t) \rangle 
\propto t^{-d/2}\,$ at long times.
In passing, it is instructive to check the case $p=1$.
It is well known that for an hypercubic lattice the DOS of
large wavelength  excitations behaves as 
$\omega (\lambda) \sim \frac{ \Omega_d}{2\,(2 \pi)^d} \:
\lambda^{d/2\, -1}\,$, where $\Omega_d$
is the surface of the $d$-dimensional unit sphere.
Thus after integrating (\ref{auto4}), we recover precisely both
the asymptotic diffusive tails and amplitudes already obtained
in Eq.\,(\ref{asympt}).

At $p = p_c$, the Lifshitz tail is again present in $\omega (\lambda)$
and swamped by a power tail which in this case has a non-trivial exponent
$d_s - 1\,$ where $d_s\,$ is, by definition, the spectral dimension
and is the ratio of the fractal dimension $d_f\,$ and the dynamic
exponent $z\,$ (which are respectively the length scaling dimensions
for mass and frequency): $d_s = d_f/z\,$. These exponents have been
investigated for the linear spin wave problem at the percolation threshold,
by direct numerical calculation \cite{Lewis} and via scaling relations 
\cite{Aharony, Houches} $d_f = d - \beta/\nu,\; z = 2 + \frac{t-\beta}{\nu}\,$
to the exponents for percolation conductance ($t\,$), correlation length
($\nu\,$) and density ($\beta\,$). For all $d > 1\,$, $d_s$ is very close
to (though not exactly equal to) the conjectured Alexander-Orbach 
\cite{Alex3} value $2/3$ (a factor of $2$ arises here because the
basic equation for ferromagnetic spin waves involves frequency squared).

Hence from Eq.\,(\ref{auto4}) we immediately obtain for the exact
finite-density autocorrelation function of hard core diffusing particles
at large times at the percolation threshold ($p=p_c\,$)
\begin{equation}
\label{robin2}
\langle C(t) \rangle \propto t^{-d_s}\,,
\end{equation}
where $d_s = (d \nu - \beta)/(2 \nu + t - \beta)\,$ is very
close to $2/3$ for all $d > 1\,$.

This is in fact the $(t/\xi^z) \to 0\,$ scaling limit of the 
following general form implied by scaling considerations applicable
when both $t$ and $\xi$ are large
\begin{equation}
\label{robin3}
\langle C(\xi, t) \rangle \propto t^{-d_s} f(t/\xi^z)\,.
\end{equation}
This form includes the crossover at $t \sim \xi^z\,$, and leads
us to conjecture that at $p_c\,$ {\it within} the percolation cluster, 
the diffusive behavior is characterized by an anomalous root mean 
square displacement $R \propto  t^{1/z}$\,, 
$z \sim  3 \,d_f/2\,$ where $d_f$ is the fractal dimension $d - \beta \nu\,$.
Eqs.(\ref{robin2}) and (\ref{robin3}) are the main results of this subsection.

\section{Conclusions}

The mapping to quantum spin systems exploited in this paper
gives a powerful approach to the stochastic dynamics of
hard core particle systems. 
Using it we have been able to obtain
in Section 2 exact dynamical properties (including the space and
time dependent pair correlation function) for the interacting system
at any density in terms of corresponding properties for linearized
spin wave dynamics.
The procedure applies equally well to disordered systems, and in addition
Goldstone arguments still apply there. In the applications made here
to disordered systems we have chosen to emphasize the diluted case,
because of the dramatic effects seen there by virtue of the underlying
percolation geometry. 

A somewhat surprising result is that hard core particle diffusion
on diluted chains is quite interesting, any dilution disconnects
the chain, so the static diffusion constant strictly vanishes. Yet
finite-time dynamics is non trivial, exhibiting scaling behavior and
stretched exponential autocorrelation functions in the time domain
(Section 3 A). The scaling behavior is analogous to that calculated
earlier for the apparently much simpler (but equivalent) problem
of ferromagnetic spin wave dynamics of diluted chains \cite{Maggs}.
There the calculations were carried out in the frequency - wave vector 
domain, for comparison with inelastic neutron scattering measurements,
so the stretched exponential time behavior was not identified,
but it should also be present there.

For interacting particle dynamics on higher dimensional diluted lattices,
by virtue of the mapping we have been able to draw on scaling procedures
and results for spin wave dynamics of diluted systems. This is 
particularly direct for the autocorrelation function of the diffusing
hard core particle system, which is given completely in terms of a DOS
which is the same as for spin wave dynamics.

Below the percolation threshold the diffusion is entirely on
finite clusters. The low frequency density of states determining
the long time behavior comes from large clusters whose occurrence
probability is exponentially small in cluster size, very much as
in the one dimensional case.
Above the percolation threshold the finite cluster contribution 
is similar, but it is now swamped by the infinite (spanning) cluster
contribution which, well away from the threshold gives a behavior
like that for the pure system. At and near the threshold ($ p \sim p_c\,$)
the diverging scale of the percolation geometry induces scaling
behavior for the long time dynamics. Here the DOS has an anomalous
power law dependence on frequency $\lambda$; the associated exponent
has been obtained via the relationship of the DOS to that for the
linear spin wave system and further relationships to percolation
processes including (static linear) conductance; the exponent in the
DOS power law is given in terms of $t,\, \beta,\, \nu\,$, the exponents
dependence on $\vert p - p_c \vert\,$ of percolation conductance,
density and correlation length. The consequent exponent in the power
law time decay of the autocorrelation function at $p_c\,$ turns out 
to be close to $2/3$ for any $d > 1\,$ (in accord with the approximate
Alexander-Orbach conjecture \cite{Alex3}\,). 
The scaling behavior of the space and
time dependent correlation function depends on the DOS exponent and also
on the length scaling exponent for the frequency $\lambda\,$. This so
called dynamic critical exponent is also related to $t,\, \beta\, \nu\,$,
and hence we can give the power law dependence of the characteristic
diffusion length on time. This generalization also allows discussion
of the crossover due to the competition between characteristic diffusion
length and percolation correlation length near the transition.

Clearly, further extensions are desirable. For the dilute systems the
most obvious ones are the calculation of space and time dependent
correlation and crossover functions, and also probability distributions
rather than just averages. In addition simulation confirmation of the
analytic predictions or experimental comparisons would be desirable.
As regards more general disorder, other types of substitutional disorder
could clearly be treated by the methods we have used. A more challenging
extension is to the biased diffusion of hard core particles: much
has been done exploiting the quantum spin mapping for the biased
case along with adsorption-desorption processes \cite{we1}. 
But biased hard core particle diffusion on disordered lattices
is at present beyond the exact techniques exploited here.

\section*{Acknowledgments}

We should like to thank D. C. Cabra for valuable discussions and
remarks. M.D.G. acknowledges financial support of CONICET, Argentina.
The research of R.B.S.  is partly supported by the EPSRC under the
Oxford Condensed Matter Theory Rolling Grant number GR/M 04426.

\newpage

\vspace{-1.5cm}
\begin{figure}
\hbox{%
\epsfxsize=4.1in
\epsffile{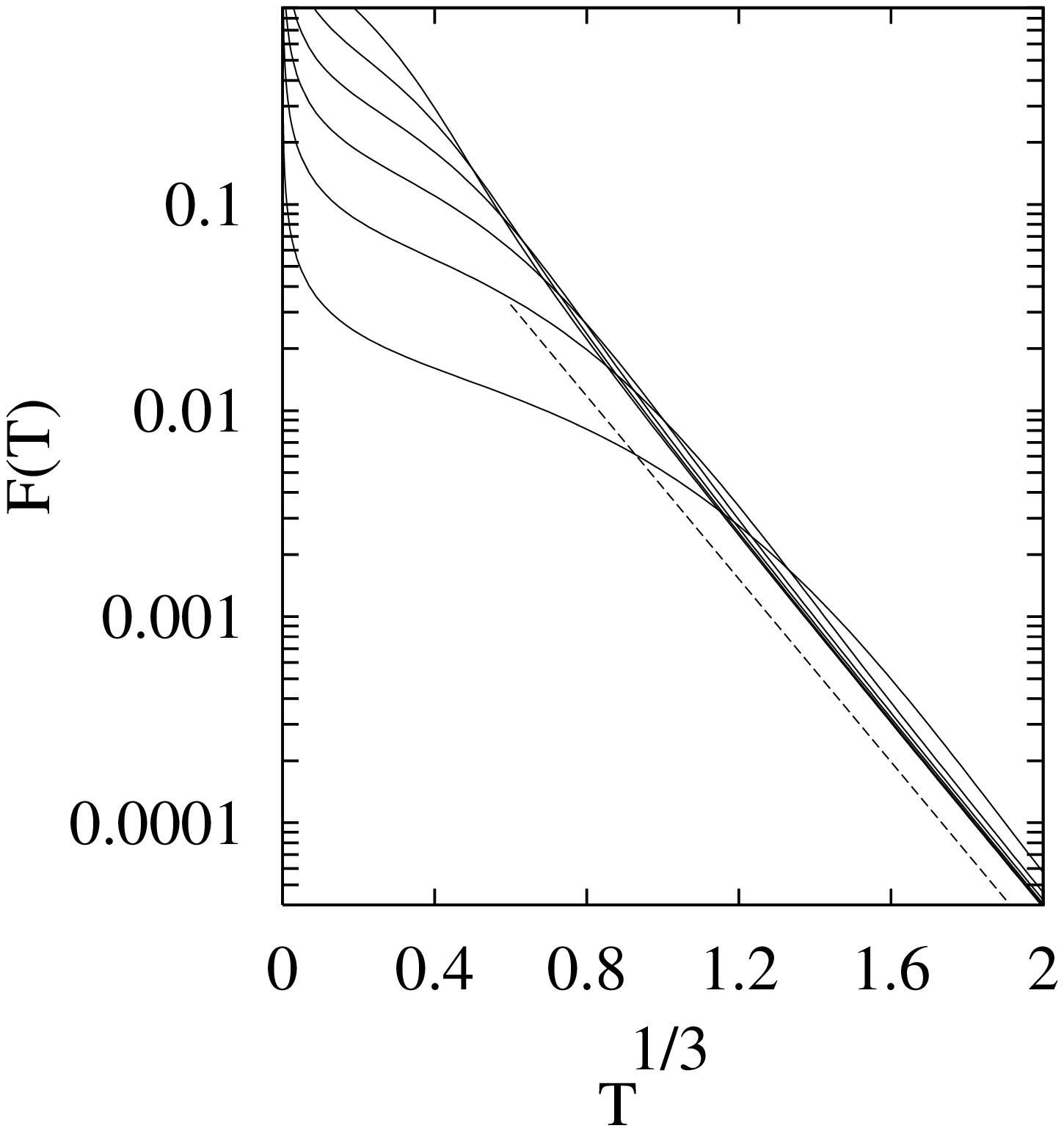}}
\vspace{1.5cm}
\caption{Stretched exponential decay and scaling regime for
autocorrelation functions of diluted chains. $\tau$ (here denoted
as $T$), is the time $t$ divided by the square of the percolation
correlation length $\xi\,$. Following
Eq.\,(38) in the text,  $F(\tau)$ is here taken as
$\langle C(\xi,\,\tau)\rangle/[\,\xi^{-1}\; \tau^{1/6}\,
\sqrt {\frac{2 a \pi}{3}}\,]$. 
Solid lines denote different degrees of dilution $p = e^{-1/\xi}\,$.
The uppermost curve at the left corresponds to $p=0.6$, and the lower curves
to $p =$ 0.5, 0.4, 0.3, 0.2 and 0.1, in descending order. The slope of the
dashed line is $-\frac{3}{2} (2 \pi)^{2/3}\;\log_{10}\, e $.}
\end{figure}


\begin{thebibliography}{100}
\bibitem{steps} For finite lattices $\Delta t \propto 1/N$, thus
late asymptotic regimes require $n \gg N$.
\end{thebibliography}
\end{document}